\documentclass[a4paper]{jpconf}
\usepackage{graphicx}
\begin{document}
\title{Effect of carrier concentration on magnetism and magnetic order in the pyrochlore iridates}

\author{M. J. Graf$^{ 1}$, S. M. Disseler$^{2}$, Chetan Dhital$^1$, T. Hogan$^1$, M. Bojko$^1$, A. Amato$^3$, H. Luetkens$^3$, C. Baines$^3$, D. Margineda$^4$, S. R. Giblin$^4$, M. Jura$^5$, and Stephen D. Wilson$^1$}

\address{$^1$ Department of Physics, Boston College, Chestnut Hill, MA 02467, USA}
\address{$^2$ Center for Neutron Research, National Institute for Standards and Technology, Gaithersburg, MD 20899 USA}
\address{$^3$ Laboratory for Muon Spin Spectroscopy, Paul Scherrer Institut, CH-5232 Villigen PSI, Switzerland}
\address{$^4$ School of Physics and Astronomy, Cardiff University, Cardiff CF24 3AA, UK}
\address{$^5$ Rutherford Appleton Laboratory, Didcot, Oxfordshire OX11 0QX, UK}

\ead{grafm@bc.edu}

\begin{abstract}
We present resistivity, magnetization, and zero field muon spin relaxation ($\mu$SR) data for the pyrochlore iridate materials Nd$_{2-x}$Ca$_{x}$Ir$_{2}$O$_{7}$ ($x = 0, 0.06$, and $0.10$) and Sm$_2$Ir$_2$O$_7$. While Nd$_{2}$Ir$_{2}$O$_{7}$ (Nd227) is weakly conducting, Sm$_{2}$Ir$_{2}$O$_{7}$ (Sm227) has slowly diverging resistivity at low temperature. Nd227 and Sm227 exhibit magnetic anomalies at $T_{M} = 105 K$ and $137 K$, respectively. However, zero-field $\mu$SR measurements show that long-range magnetic order of the Ir$^{4+}$ sublattice sets in at much lower temperatures ($T_{LRO} \sim 8 K$ for Nd227 and $70 K$ for Sm227); both materials show heavily damped muon precession with a characteristic frequency near 9 MHz. The magnetic anomaly at $T_{M}$ in Nd227 is not significantly affected by the introduction of hole carriers by Ca-substitution in the conducting Nd$_{2-x}$Ca$_{x}$Ir$_{2}$O$_{7}$ samples, but the muon precession is fully suppressed for both. 
\end{abstract}

\section{Introduction}
Strongly correlated transition metal oxides exhibit a wide-range of important and interesting phenomena, including high temperature superconductivity and colossal magnetoresistance. While iridium oxides have a much reduced Ir-Ir correlation energy due to their extended 5d orbitals, the strong spin orbit interaction (SOI) energy makes possible a SOI-assisted Mott-like insulating state with $J_{eff} = 1/2$, as observed in Sr$_2$IrO$_4$ [1]. The pyrochlore iridates A$_{2}$Ir$_{2}$O$_{7}$  (A = lanthanide or Y), or A227, consist of interpenetrating corner-sharing tetragonal structures for both the A$^{3+}$ and Ir$^{4+}$ ions. This structure suggests that frustration may also play a role in determining the magnetic ground state. For all members of the family (apart from A = Pr), a transition in the DC magnetic susceptibility associated with the Ir sublattice is observed at a temperature $T_{M}$, which ranges from near 200 K (A = Y) to near 30 K (A = Nd). The deviation of field-cooled (FC) from zero-field-cooled (ZFC) data below $T_{M}$ bears a resemblance to a spin-glass transition. For small A-ion radii $R_{A}$ an insulating low temperature ground state is observed, and for several of species of A a concomitant metal-insulator transition has been reported at temperature $T_{MI} \sim T_{M}$ [2]. Zero field muon spin rotation ($\mu$SR) measurements confirm long-range magnetic ordering of the Ir sublattice at temperatures $T_{LRO} = 150 K, 180 K$ and $135 K$ for the insulators Yb227, Y227 [3], and Eu227 [4], respectively. The muon precession frequencies are near 14 MHz, corresponding to an Ir$^{4+}$ moment $\mu_{Ir} \sim 0.3 \mu_{B}$ [5], and are lightly damped. The moments order in an all-in/all-out (AIAO) magnetic structure, where the Ir$^{4+}$ moments are pointing either all in towards, or all away from, the center of a given tetrahedron, as indicated by  $\mu$SR measurements [3] and analysis [5] and resonant inelastic x-ray scattering measurements [6].

In contrast, Pr227, with a large $R_{A}$, has a conducting ground state and shows no magnetic transition down to $T = 25 mK$ [7]. This system has been proposed to be a chiral spin liquid at low temperatures [8]. Factors that may play a role in the crossover from magnetically ordered insulator to a conductor lacking long-range order include the A-ion radius and magnetic moment, the carrier density, geometric frustration effects, and structural and stoichiometric disorder. This crossover region, i.e., near A $=$ Nd and Sm, is of particular interest. Recent theoretical work indicates the possible existence of topologically non-trivial ground states in this region that may require the presence of magnetic order to break time-reversal symmetry [9]. The Nd227 ground state appears to be very sensitive to the details in preparation, as expected in a crossover region: samples of comparable purity and quality as determined by x-ray measurements can be either insulating or weakly conducting. The weakly conducting samples have a relatively high $T_{M} = 105 K$ [10] compared to the insulating samples with $T_{MI} \sim T_{M} = 33 K$ [2]. However, regardless of whether the low temperature state is insulating or conducting, long range magnetic order in the Ir$^{4+}$ sublattice is observed in Nd227 with a muon precession frequency of roughly 9 MHz and with very strong damping [10,11]. For the insulating samples $T_{LRO} \sim T_{M} = 33 K$, while for the weakly conducting samples $T_{LRO} \sim 8 K \ll T_{M}$, suggesting an extended region of a magnetic phase with short-range order that appears to be correlated with the presence of additional charge carriers. The magnetic state of the Nd$^{3+}$ ions remains an open question: neutron scattering measurements indicate the Nd-sublattice also orders in an AIAO structure [12], but combined $\mu$SR [10] and Hall and magnetization measurements [13] suggest a spin-ice like state.

In this work we present preliminary results on the effects of carrier density on the the magnetic transitions at temperatures $T_{M}$ and $T_{LRO}$ in Nd227 by introducing hole carriers through Ca-substitution. Additionally, we present the first $\mu$SR measurements on Sm227. 

\section{Samples}
Polycrystalline samples of Nd227 (Sm227) were synthesized by reacting stoichiometric amounts of 99.99$\%$ purity Nd$_2$O$_3$ (Sm$_2$O$_3$) and IrO$_2$ (99.9$\%$) in air to minimize variations in the oxygen stoichiometry. Powders were pelletized using an isostatic cold press and reacted between 900 C to 1125 C over six days with several intermediate grindings. The samples were determined to be phase-pure with the exception of two minor impurity phases of IrO$_2$ and Nd$_2$O$_3$ (Sm$_2$O$_3$) each comprising less than 1$\%$ of the total volume fraction. Powder x-ray diffraction measurements yielded narrow lineshapes comparable to the best previously reported data on Nd227 and Sm227, demonstrating the high homogeneity of the sample stoichiometry. Nd$_{2-x}$Ca$_{x}$Ir$_{2}$O$_{7}$ polycrystalline samples with $x = 0.06$ and $0.10$ were prepared by reacting stoichiometric amounts of Nd$_2$O$_3$, CaCo$_3$ and IrO$_2$ in air as described above. Several EDS measurements taken on the nominal $x = 0.10$ sample over areas of 100 $\mu m^2$ show that the average Ca concentration was $x = 0.08$. X-ray refinement yields a slight decrease in the lattice parameter from $a_{Nd} = 10.375$ to $10.351 \AA$ for the nominal $x = 0.10$ sample.

\section{Resistivity}
The temperature dependent resistivity was measured in the range $0.3 < T < 300 K$ via a four-point AC resistance bridge. The samples were pressed sintered pellets, with an error in the resistivity of approximately 20$\%$ based on uncertainties in the sample dimensions. Amplitude dependence studies at the lowest temperatures ensured there was no self heating of the samples; we note that Sm227 has a very long thermal equibration time. In Figure 1, we show the temperature dependent resistivity for Sm227 and Nd227, along with previously published data for Yb227 [3]. Yb227 is insulating and exhibits a temperature dependence consistent with variable range hopping at low temperatures. Nd227 exhibits a Kondo-like behavior, with a logarithmic upturn below 10 K and saturation at very low temperatures.  Sm227 exhibits a behavior intermediate between Yb227 and Nd227, and is qualitatively similar to earlier results [2], with a crossover from high temperature conducting behavior to an insulating behavior at low temperatures. The crossover from conducting to insulating behavior occurs at $T \sim 70 K < T_{M}$ (see next section), with a resistivity that is slowly divergent down to 0.3 K. This does not follow the variable range hopping law, but rather following an inverse power law $T^{-\beta}$ with $\beta = 0.52(2)$.

Also shown in Figure 1 are data for Nd$_{1.94}$Ca$_{0.06}$Ir$_2$O$_7$ and Nd$_{1.9}$Ca$_{0.1}$Ir$_2$O$_7$. Assuming one hole carrier per Ca ion, the expected hole carrier density in the $x = 0.06$ sample should be approximately twice the estimated upper limit for the undoped Nd227 carrier density from Hall effect measurements [13]; Hall measurements are now underway to confirm the actual carrier densities. We observe a reduction in both the overall value of the resistivity and in the low temperature upturn relative to the room temperature resistivity, with the unexpected result that the room temperature resistivity of $x = 0.06$ being lower than that of the $x = 0.10$ sample. The unusual behavior is illustrated in the inset to Figure 1, where the fractional change of resistance (relative to room temperature) versus temperature is shown. At present it is not clear if this is due to intrinsic effects, such as enhanced scattering by Ca impurities, or extrinsic effects (e.g. subtle differences in pressure or temperature while pressing the powders).

\begin{figure}[h]
\begin{minipage}{18pc}
\includegraphics[width=18pc]{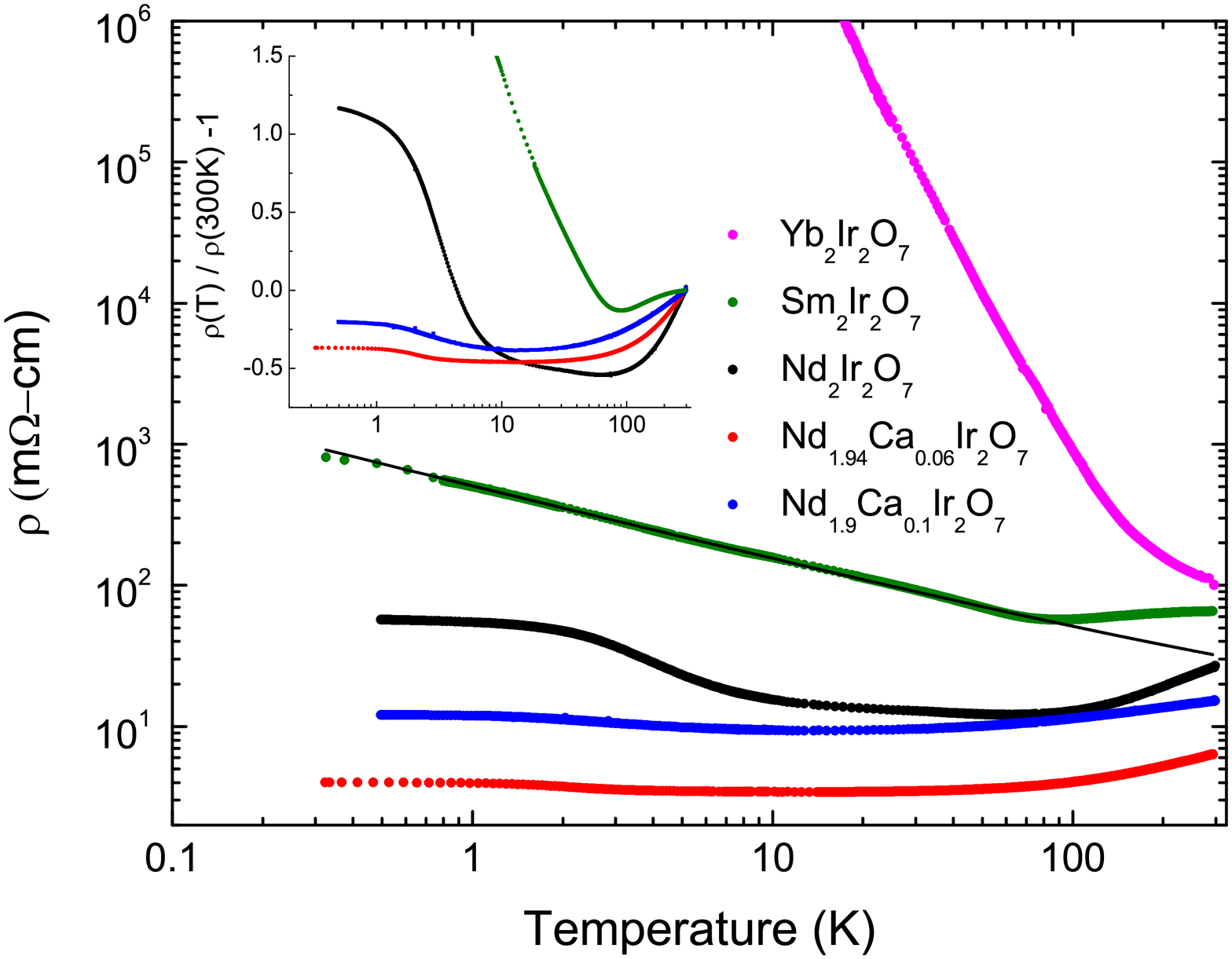}
\caption{\label{label}The temperature dependent resistivity for several pyrochlore iridates; data for Yb227 is taken from Ref. 3. The solid line is a fit to the Sm227 data at low temperatures (see text).}
\end{minipage}\hspace{1pc}%
\begin{minipage}{18pc}
\includegraphics[width=18pc]{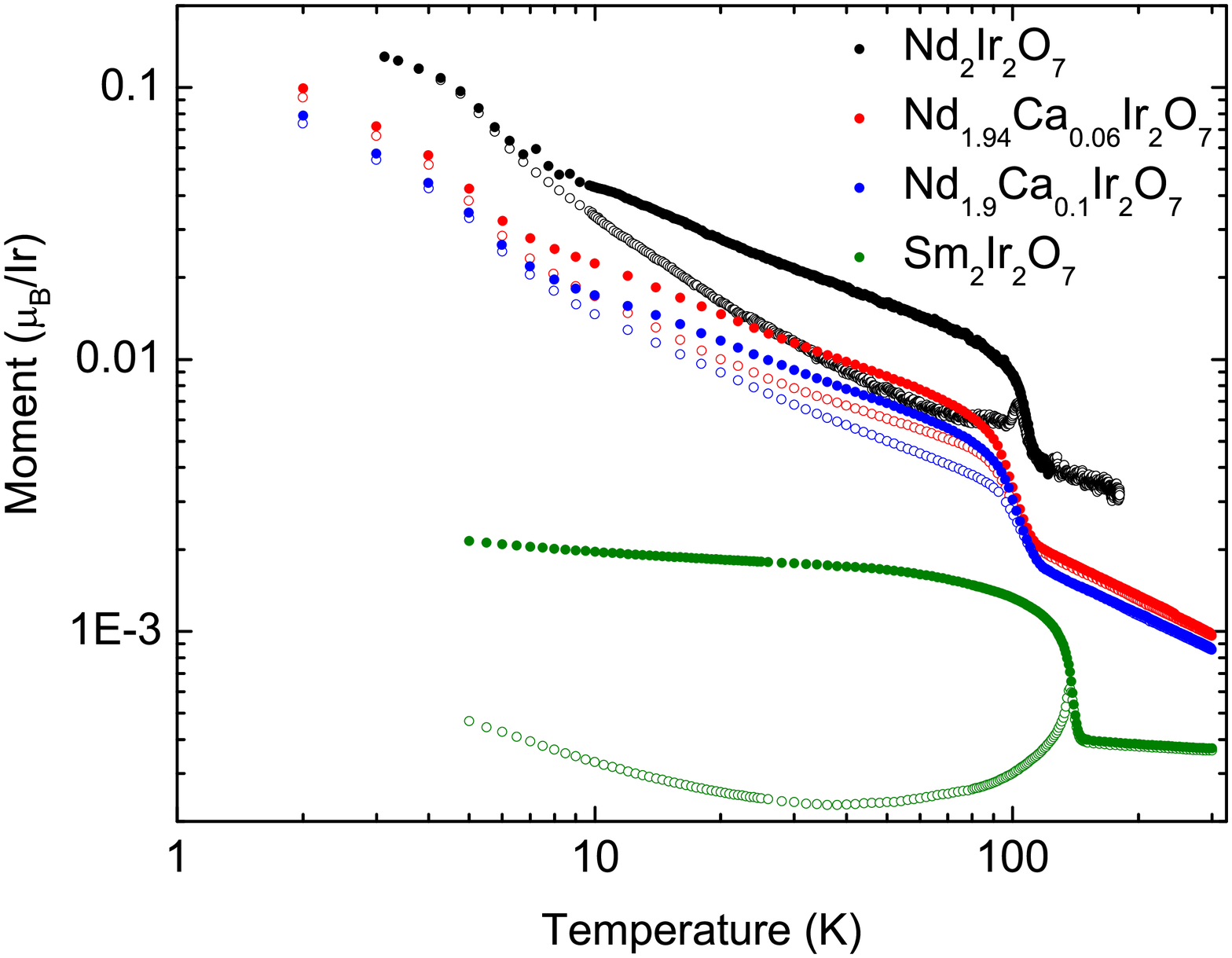}
\caption{\label{label}The temperature dependent magnetization in fields of 1000 Oe for several pyrochlore iridates (open circles: zero field cooled; filled circles: field cooled).}
\end{minipage} 
\end{figure}

\section{Magnetization}
Data were taken in Quantum Design MPMS SQUID magnetometers located in Cardiff and Didcot. In Figure 2 we show the temperature-dependent DC magnetization for the Nd-based samples and Sm227 (taken in fields of 1000 Oe). It is immediately evident that $T_{M} = 105 K$ is not significantly modifed by the Ca-substitution. Data for Sm227 shows that $T_{M} = 137 K$, higher than the value 118 K observed in Ref. 2. The Sm$^{3+}$ contribution to the observed magnetization is significantly smaller than that of Nd$^{3+}$. Assuming that the ordered Ir$^{4+}$ sublattice makes a nearly temperature independent contribution to the DC susceptibility at low temperature, we fit the zero-field cooled data below 20 K to a Curie-Weiss form including an additive constant. This yields an effective moment of $\mu_{Sm} = 0.28(1) \mu_B$ and a Curie-Weiss temperature of $T_{CW} = -0.9(1) K$, indicative of weak antiferromagnetic correlations. This value of $\mu_{Sm}$ is well below the free ion value $0.8 \mu_B$, but not unusual,  as seen for Sm$_2$Ti$_2$O$_7$ [14], which has a very small g-factor [15].

\section{Zero-field $\mu$SR}
Data were taken at PSI using the GPS and LTF spectrometers and at ISIS on the EMU spectrometer. In Figure 3 we show zero-field muon depolarization curves at $T = 25 mK$ for the Nd-based and Sm227 samples. Also shown for comparison is the depolarization of insulating Yb227 [3]; the Yb$^{3+}$ has a large moment ($\mu_{Yb} = 3.3 \mu_B$) which causes damping of the muon precession at low temperatures. Three points are immediately obvious: (1) the depolarization in the Nd227 and Sm227 are nearly identical; (2) these are both at significantly lower frequency and are more strongly damped than oscillations observed in insulating A227 materials (e.g., Yb227), and (3) there are no signs of muon precession in the Ca-substituted samples.   

\begin{figure}[h]
\begin{minipage}{18pc}
\includegraphics[width=18pc]{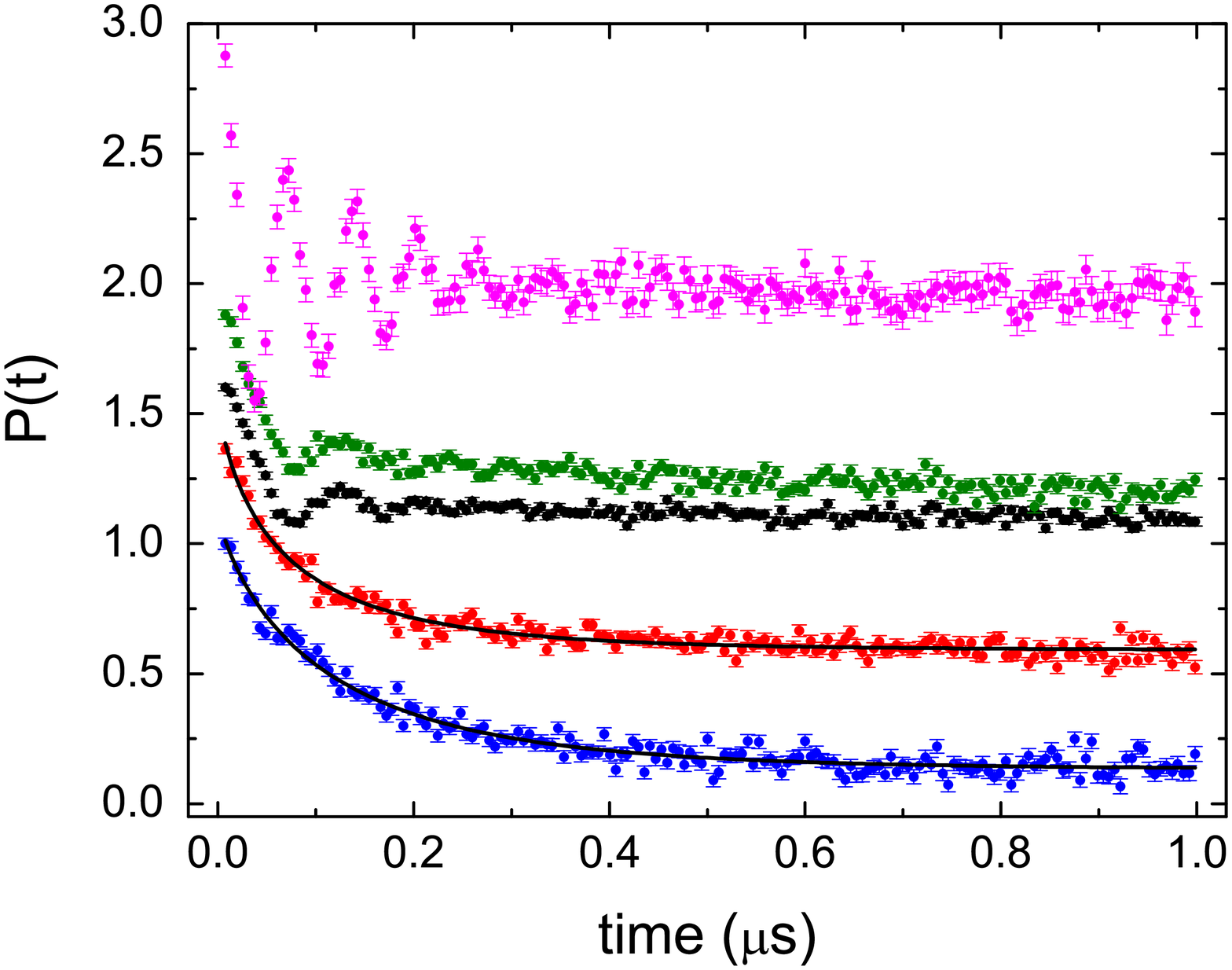}
\caption{\label{abel}Zero field depolarization with time at $T = 25 mK$. From top to bottom: Yb227, Sm227, Nd227, Nd$_{1.94}$Ca$_{0.06}$Ir$_{2}$O$_{7}$, and Nd$_{1.9}$Ca$_{0.1}$Ir$_{2}$O$_{7}$. Curves are offset by constant amounts (except for the Yb227 curve due to the large oscillations). Solid lines are fits as described in the text.}
\end{minipage}\hspace{1pc}%
\begin{minipage}{18pc}
\includegraphics[width=18pc]{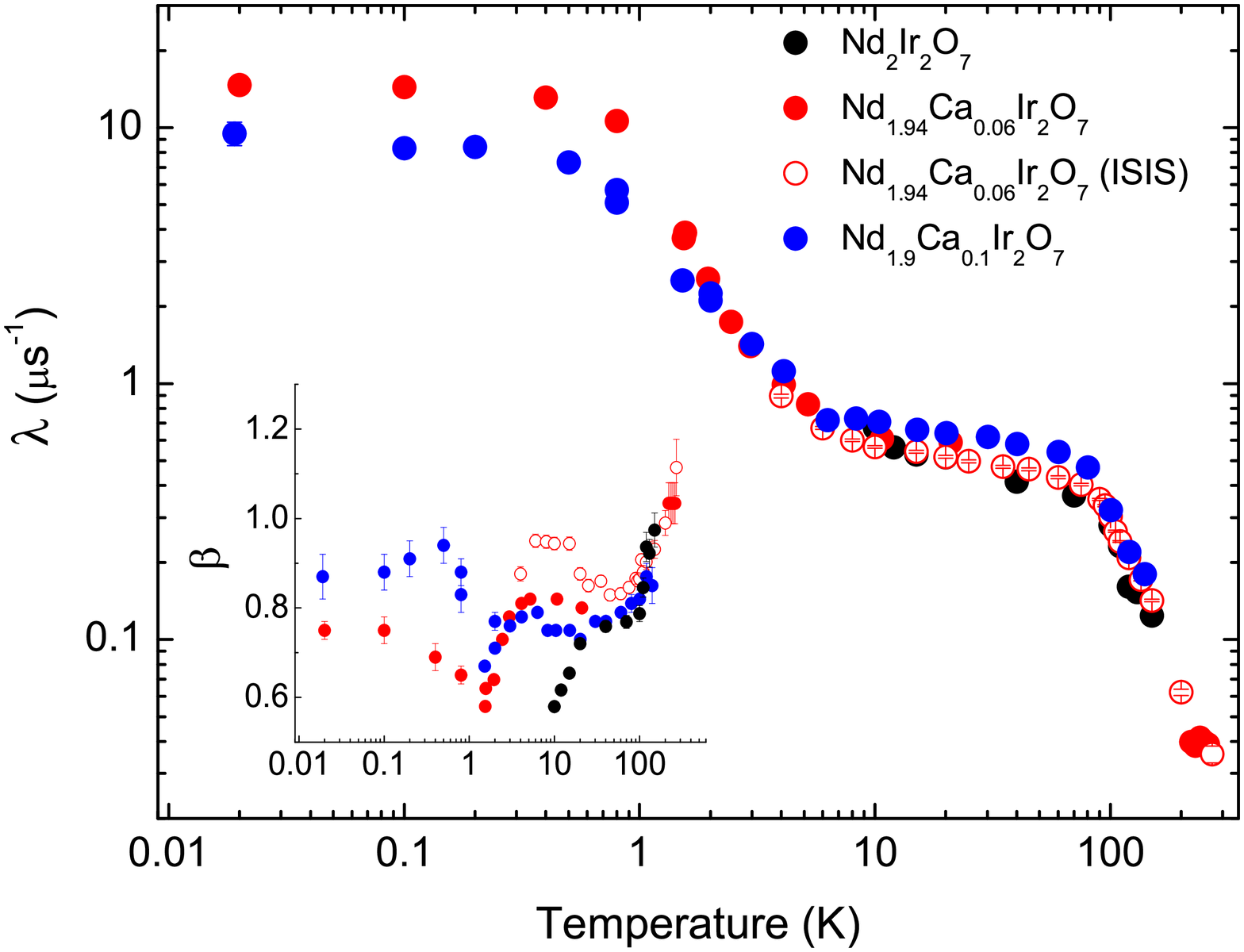}
\caption{\label{label}Temperature dependence of the stretched exponential depolarization rate for Nd227, Nd$_{1.94}$Ca$_{0.06}$Ir$_{2}$O$_{7}$, and Nd$_{1.9}$Ca$_{0.1}$Ir$_{2}$O$_{7}$. Inset: temperature dependence of the exponent.}
\end{minipage} 
\end{figure}

The strikingly different depolarization of Sm227 and Nd227 compared to that of the insulating A227 materials is not yet understood, and may be due to reduced Ir$^{4+}$ moment [15], a different magnetic ordering structure, a different muon stopping site, or the presence of magnetic A-ions. The last possibility is unlikely, given that the Yb$^{3+}$ has a larger moment than both the Nd$^{3+}$ ($\mu_{Nd} = 2.8 \mu_B$) or the Sm$^{3+}$. The muon stopping site is also unlikely to be different: computer calculations [6, 10] have been made for the electrostatic potential in Sm227, and yield nearly identical results as for Y227. A reduction in the Ir moment is an appealing scenario, but is hard to reconcile with the very small change in lattice parameters concomitant with the dramatic change in magnetic behavior between Eu227 and Sm227. Recent reanalysis of the Nd227 $\mu$SR results shows that a Bessel depolarization function yields a superior fit than the phenomenological function used in Ref. 10, consistent with an incommensurate, rather than AIAO, ordering of the Ir$^{4+}$ sublattice for both Nd227 and Sm227. This leads us to believe that the delicately balanced magnetic interactions shift between Eu227 and Sm227, producing a different (and possibly incommensurate) ordering structure. A detailed analysis will be presented in a future publication. The change in magnetic structure is not solely determined by carrier density, since both insulating and weakly conducting samples of Nd227 have similar muon precession frequencies and very strong damping [10, 11]. 

From the Ca-substituted data we infer that increasing the carrier density strongly suppresses the onset of magnetic ordering simultaneous to destabilizing the spin-orbit Mott phase. This observation is consistent with the earlier observation that while insulating Nd227 shows spontaneous muon precession near 33 K, for weakly conducting Nd227 the onset temperature is roughly 8 K. In contrast, the carrier density has little effect on the magnetic transition at $T_{M}$, as shown in Figure 2. The Ca-substituted samples have been fit to a stretched exponential depolarization function (solid lines in Figure 3), and the temperature dependences of the depolarization rates and exponents are shown in Figure 4. Also shown is the stretched exponential fit for Nd227 for $T \geq 10 K > T_{LRO}$. We associate the rise in the depolarization rate below 10 K with freezing of the Nd$^{3+}$ moments.  

\begin{figure}[h]
\begin{minipage}{18pc}
\includegraphics[width=18pc]{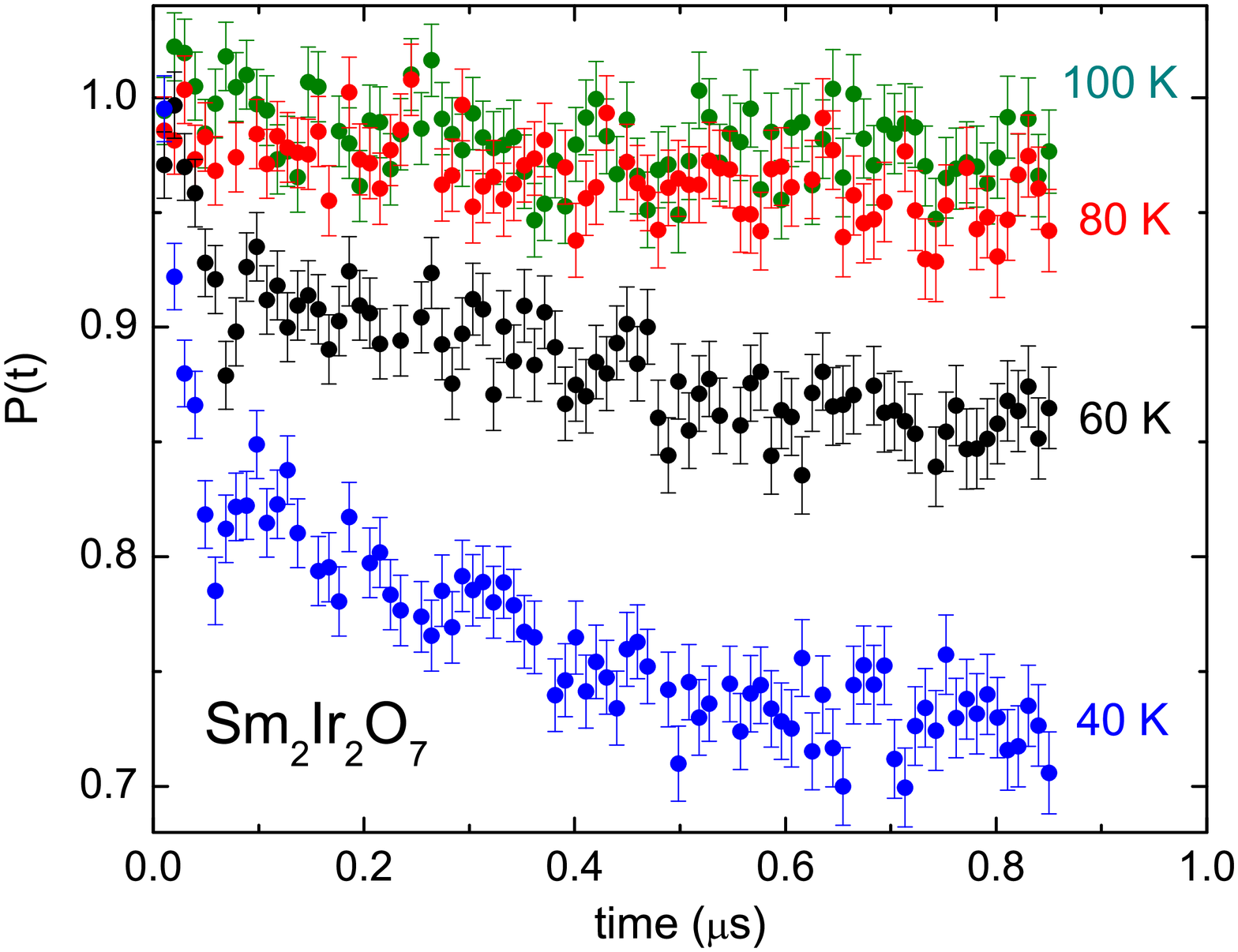}
\caption{\label{abel}Evolution of the zero field depolarization of Sm227 with temperature in the vicinity of $T_{LRO}$.}
\end{minipage}\hspace{1pc}%
\begin{minipage}{18pc}
\includegraphics[width=18pc]{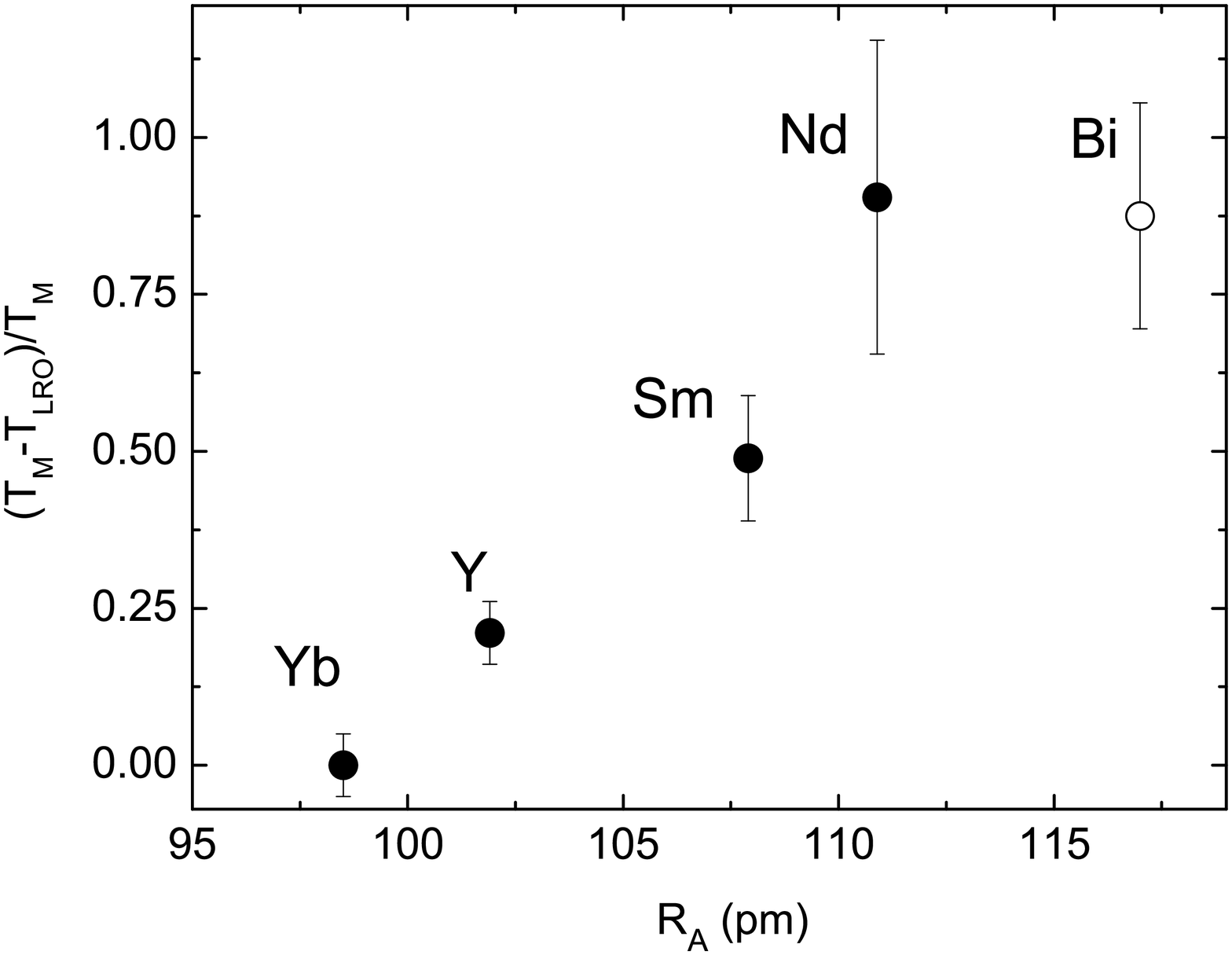}
\caption{\label{label}The variation of the fractional suppression of $T_{LRO}$ with respect to  $T_M$ with A-ion radius. Bi227 data taken from Ref. 16.}
\end{minipage} 
\end{figure}

Tracking the temperature dependence of the Sm227 ZF-$\mu$SR depolarization shows that an oscillatory-like time dependence sets in below $T = 80 K$, as shown in Figure 5. Based solely on this data we cannot determine if this is connected to true long-range magnetic ordering in the Ir sublattice, or simply due to Ir-moment spin freezing. However, taken in context of the Nd227 measurements for which the oscillatory behavior is more clearly distinguished from Kubo-Gauss depolarization [10], we assume that it is indeed due to the onset of long-range magnetic order, and so for Sm227 $T_{LRO} \sim 70 K$. This is well below $T_{M} = 137 K$, and is consistent with our trend reported earlier: as the A-ion radius is increased a short range ordered magnetic phase is stabilized such that $T_{LRO} < T_{M}$. In Figure 6 we have plotted the fractional suppression $1-T_{LRO} / T_{M}$ versus A-ion radius, as extracted from our data for Yb227, Y227, Sm227, and Nd227; also included is the fractional difference in the two magnetic transition temperatures reported for Bi227 [16], although the exact nature of those two transitions is not clear. The correlation is evident, though the underlying mechanism remains unclear.

\section{Summary and Conclusions}
Addition of carriers to Nd227 via Ca substitution has little effect on the magnetization anomaly observed at $T_{M}$, but completely suppresses the onset of long-range magnetic order; it would be of interest to study the effects of Ca-substitution on $T_{M}$ for insulating samples of Nd227 and other insulating pyrochlore iridates as in Ref. 17. Additionally, Sm227 has magnetic properties that are very similar to Nd227: $T_{LRO} < T_{M}$ and heavily damped muon precession near 9 MHz. The underlying mechanisms causing the stabilization of a magnetic phase lacking long-range order and the abrupt change in magnetic ordering between Eu227 and Sm227 remain unknown. Detailed studies of the changes in stoichiometry [18,19] and possible distortions of the local A-ion environment with A-ion radius and preparation technique would appear to be necessary to fully understand these behaviors.

\section{Acknowledgments}
$\mu$SR measurements were performed at the Swiss Muon Source and the ISIS Muon Facility. This work was supported in part by the National Science Foundation (SDW and MJG), the National Research Council (SMD), and EPSCR (SRG).


\section*{References}
\begin{thebibliography}{9}
\bibitem{Ref1} B. J. Kim, H. Jin, S. J. Moon, J.-Y. Kim, B.-G. Park, C. S. Leem, J. Yu, T. W. Noh, C. Kim, S.-J. Oh, J. –H. Park, V. Durairaj, G. Cao, and E. Rotenberg 2008 {\it Phys. Rev. Lett.} {\bf 101} 076402
 \bibitem{Ref2} K. Matsuhira, M. Wakeshima, R. Nakanishi, T. Yamada, A. Nakamura, W. Kawano, S. Takagi, and Y. Hinatsu 2007 {\it J. Phys. Soc. Jpn.} {\bf 76} 04370 ; K. Matsuhira, M. Wakeshima, Y. Hinatsu, and S. Takagi 2011 {\it J. Phys. Soc. Jpn.} {\bf 80} 094701 
 \bibitem{Ref3} S. M. Disseler, C. Dhital, A. Amato, S. R. Giblin, C. de la Cruz, S. D. Wilson, and M. J. Graf 2012 {\it Phys. Rev. B} {\bf 86} 014428 
 \bibitem{Ref4} S. Zhao, J. M. Mackie, D. E. MacLaughlin, O. O. Bernal, J. J. Ishikawa, Y. Ohta, and S. Nakatsuji 2011 {\it Phys. Rev. B} {\bf 83} 180402R  
 \bibitem{Ref5} S. M. Disseler, {\it Phys. Rev. B} {\bf 89} 140413R
 \bibitem{Ref6} H. Sagayama, D. Uematsu, T. Arima, K. Sugimoto, J. J. Ishikawa, E. O’Farrell, and S. Nakatsuji 2013 {\it Phys. Rev. B} {\bf 87} 100403R
 \bibitem{Ref7} D. E. MacLaughlin, Y. Ohta, Y. Machida, S. Nakatsuji, G. M. Luke, K. Ishida, R. H. Heffner, L. Shu, and O. O. Bernal 2008 {\it Physica B} {\bf 404} 667 
 \bibitem{Ref8} Y. Machida, S. Nakatsuji, S. Onoda, T. Tayama, and T. Sakakibara 2010 {\it Nature} {\bf463} 210 
 \bibitem{Ref9} W. Witczak-Krempa, G. Chen, Y. B. Kim, and L. Balents 2014 {\it Annu. Rev. Condens. Matter Phys.} {\bf 5} 57 
 \bibitem{Ref10} S. M. Disseler, C. Dhital, T. C. Hogan, A. Amato, S. R. Giblin, C. de la Cruz, A. Daoud-Aladine, S. D. Wilson, and M. J. Graf 2012 {\it Phys. Rev. B} {\bf 85} 174441 
 \bibitem{Re11} H. Guo, K. Matsuhira, I. Kawasaki, M. Wakeshima, Y. Hinatsu, I. Watanabe, and Z. Xu 2013 {\it Phys. Rev. B} {\bf 88} 060411 
 \bibitem{Ref12} K. Tomiyasu, K. Matsuhira, K. Iwasa, M. Watahiki, S. Takagi, M. Wakeshima, Y. Hinatsu, M. Yokoyama, K. Ohoyama, and K.Yamada 2012 {\it J. Phys. Soc. Jpn.} {\bf 81} 034709
 \bibitem{Ref13} S. M. Disseler, S. R. Giblin, C. Dhital, K. C. Lukas, S. D. Wilson, and M. J. Graf 2013 {\it Phys. Rev. B} {\bf 87} 060403R
 \bibitem{Ref14} S. Singh, S. Saha, S. K. Dhar, R. Suryanarayanan, A. K. Sood, and A. Revcolevschi 2008 {\it Phys. Rev. B} {\bf 77} 054408
 \bibitem{Ref15} B. Z. Malkin, T. T. A. Lummen, P. H. M. van Loosdrecht, G. Dhalenne, and A. R. Zakirov 2010 {\it J. Phys. Condens. Matter} {\bf 22} 276003
 \bibitem{Ref16} P. J. Baker, J. S. Möller, F. L. Pratt, W. Hayes, S. J. Blundell, T. Lancaster, T. F. Qi, and G. Cao 2013 {\it Phys. Rev. B} {\bf 87} 180409R
\bibitem{Ref17} H. Fukazawa and Y. Maeno 2002 {\it J. Phys. Soc. Jpn.} {\bf 71} 2578
 \bibitem{Ref18} J. J. Ishikawa, E. C. T. O’Farrell, and S. Nakatsuji 2012 {\it Phys. Rev. B} {\bf 85} 245109 
 \bibitem{Ref19} See for example Section II.C in J. S. Gardner, M. J. P. Gingras, and J. E. Greedan 2010 {\it Rev. Mod. Phys.} {\bf 82} 53 (Section II.C)
\end{thebibliography}
\end{document}